\documentclass{aa}  
\usepackage{graphicx}
\usepackage{lscape}
\usepackage[breaklinks=true]{hyperref}
\hypersetup{colorlinks=true,urlcolor=blue,citecolor=blue,pdfborder= 0 0 0}
\bibpunct{(}{)}{;}{a}{}{,} 
\usepackage{txfonts}
\usepackage{etex}
\reserveinserts{18}
\usepackage{morefloats}

\begin{document} 

   \title{The effect of cosmic web filaments on the properties of groups and their central galaxies}
   \subtitle{}
   \author{A. Poudel
          \inst{1,\star}
          \and
          P. Heinämäki\inst{1}\
          \and
          E. Tempel\inst{2,3}
       \and
           M. Einasto\inst{2}
       \and
          H. Lietzen\inst{1}         
      \and
      P. Nurmi\inst{1,4}
                   }
           
   \institute{\inst{1} Tuorla Observatory, University of Turku,
              Väisäläntie 20, Piikkiö, Finland\\
              \inst{2} Tartu Observatory, Observatooriumi 1, 61602 T\~oravere, Estonia\\
              \inst{3} Leibniz-Institut für Astrophysik Potsdam (AIP), An der Sternwarte 16, D-14482 Potsdam, Germany\\
              \inst{4} Finnish Centre for Astronomy with ESO (FINCA), University of Turku, Väisäläntie 20, Piikkiö, Finland\\
              $^{\star}$ \email{anuppou@utu.fi}
              }
   \date{Received ; accepted }
 
  \abstract
 {The nature versus nurture scenario in galaxy and group evolution is a long-standing problem not yet fully understood on cosmological scales.}   
   {We study the properties of groups and their central galaxies in different large-scale environments defined by the luminosity density field and the cosmic web filaments.
}
   { We use the luminosity density field constructed using 8~$h^{-1}$Mpc smoothing to characterize the large-scale environments. We use the Bisous model to extract the filamentary structures in different large-scale environments. We study the properties of galaxy groups as a function of their dynamical mass in different large-scale environments.}
   {We find differences in the properties of central galaxies and their groups in and outside of filaments at fixed halo and large-scale environments. In high-density environments, the group mass function has higher number densities in filaments compared to that outside of filaments towards the massive end. The relation is opposite in low-density environments. At fixed group mass and large-scale luminosity density, mass-to-light ratios show that groups in filaments are slightly more luminous than those outside of filaments. At fixed group mass and large-scale luminosity density, central galaxies in filaments have redder colors, higher stellar masses, and lower specific star formation rates than those outside of filaments. However, the differences in central galaxy and group properties in and outside of filaments are not clear in some group mass bins. We show that the differences in central galaxy properties are due to the higher abundances of elliptical galaxies in filaments. 
}
   {Filamentary structures in the cosmic web are not simply visual associations of galaxies, but rather play an important role in shaping the properties of groups and their central galaxies. The differences in central galaxy and group properties in and outside of cosmic web filaments are not simple effects related to large-scale environmental density. The results point towards an efficient mechanism in cosmic web filaments which quench star formation and transform central galaxy morphology from late to early types. 
}

   \maketitle
%
\section{Introduction}
According to the standard theory of cosmic structure formation, the currently observable cosmic web formed through gravitational enhancement of very small dark matter density fluctuations present in the very early Universe \citep[e.g.,][]{1978IAUS...79..241J,1986ApJ...302L...1D}. Over the cosmic time, when these fluctuations became large enough, they stopped growing and collapsed under gravity to form virialized regions, also called dark matter halos. Baryons then cooled, condensed, and formed galaxies in the potential wells of these dark matter halos \citep{1978MNRAS.183..341W}. Dark matter halos then grew by merging with other halos or accreting mass from its surroundings to form a wide range of structures from small groups to massive clusters. Presently, these groups and clusters are found to be organized in a web-like pattern called the cosmic web, which is dominated by filamentary structures containing almost half of observed galaxies and mass in the local Universe \citep{2014A&A...572A...8T}. The filaments act as connecting links between clusters and are separated by nearly empty voids. It is a topic of debate if these large-scale structures in the cosmic web have played any role in the evolution of galaxies and groups. 

Standard theories assume that the halo environment, where galaxies are initially formed, plays the dominant role in shaping the properties of galaxies. In support of this scenario, the mass of the halo is found to correlate with many galaxy properties like, stellar mass \citep{2010ApJ...717..379B}, specific star formation rate, color \citep{2006MNRAS.366....2W} and morphology \citep{2012ApJ...746..160W}. However, correlations between galaxy and group properties and environments have been found to exist also on scales much larger than halos \citep{2012A&A...545A.104L,2013MNRAS.432.1367L,2014A&A...562A..87E,2015MNRAS.448.1483L}. Using the luminosity density method with a smoothing scale of 8~$h^{-1}$Mpc \citep{2007A&A...464..815E,2012A&A...539A..80L}, \citet{2016A&A...590A..29P} found that the low mass end slope of the stellar mass function of satellite galaxies is steeper in high-density environments compared to low-density environments. They also found that groups with similar masses are richer in high-density environments compared to low-density environments, irrespective of the galaxy morphologies. These results contradict with several other studies which indicate that the halo occupation distribution at fixed halo mass is statistically independent of the large-scale environment \citep{1991ApJ...379..440B,1999MNRAS.302..111L,2004ApJ...609...35K}. In simulations, the turnover mass at the massive end of the dark matter halo mass functions is the highest in clusters and gradually decreases toward filaments, sheets, and voids \citep{2007MNRAS.375..489H,2015MNRAS.446.1458M}. Any correlation between surrounding large-scale structure and internal galaxy properties may simply be due to the differences in halo masses in different cosmic web environments. In this regard, it is essential to fix the halo environment to study the true effect of the large-scale structures on galaxy and group properties. Using high-resolution hydrodynamical simulations, \citet{2015ApJ...812..104T} have found that central galaxies in similar mass halos in large-scale ($\sim$20\,Mpc) overdense regions have higher stellar masses than those in underdense regions. 

Apart from large-scale galaxy densities, galaxy and group properties have been also found to correlate with filamentary structures in the cosmic web. Groups in filaments are found to have more satellites than outside of filaments \citep{2015ApJ...800..112G} and the satellites tend to align with galaxy filaments \citep{2015MNRAS.450.2727T}. Using a sample of 1,799 isolated spiral galaxies, \citet{2016MNRAS.457.2287A} found that the stellar mass of isolated spiral galaxies decreases slightly as a function of distance from the filament axis whereas the star formation rates rises slightly with distance. The spins of bright spiral galaxies tend to have parallel alignment with filaments, while elliptical and S0 galaxies have their spins aligned perpendicular to the filament direction \citep{2013MNRAS.428.1827T, 2013ApJ...775L..42T}. However, none of these studies have fixed the halo environment which is one of the main focus of this paper.

In this paper, we study the effect of cosmic web filaments on properties of groups and their central galaxies using a statistically large sample of galaxies and groups. Due to the observational biases in large galaxy surveys and unvirialized nature of the large-scale structures, the characterization of the large-scale structure is a non-trivial task. In this study, we first use the luminosity density method to define low- and high-density, large-scale environments (voids, intermediate regions, and superclusters) and then the method based on marked point processes to extract filamentary structures in these environments \citep{2016A&C....16...17T}. The outline of the paper is as follows: in sections~\ref{sec:data} and~\ref{sec:methods}, we explain the data and methods used in this study, respectively. In sections~\ref{sec:results(groups)} and~\ref{sec:results(central galaxies)}, we present the results of our analysis of the environmental influence on group and their central galaxy properties. In Section~\ref{sec:discussion}, we discuss the implications of our results on central galaxy and group evolution. Finally, we summarize our results in Section~\ref{sec:conclusion}. Throughout the paper, we adopt, $H_0$ = 100 $h$ km s$^{-1}$ Mpc$^{-1}$, $\Omega_\mathrm{m}$ = 0.27 and $\Omega_{\Lambda}$ = 0.73. 

\section{Data}
\label{sec:data}
\subsection{Group catalogs}
The group catalogs used in this work are taken from \citet{2014A&A...566A...1T} that uses the SDSS DR10 data and a friends-of-friends algorithm with a redshift dependent linking length to define groups of galaxies. The main catalog is flux limited with a magnitude limit of 17.77 mag in the SDSS $r$ band and consists of 588193 galaxies and 82458 groups. The catalog covers an area of around 7221 square degrees extending up to redshift 0.2. Seven different volume-limited catalogs of galaxies are extracted from the main flux limited sample and are complete at absolute magnitudes down to M$_{r,\mathrm{lim}}$ = $-$18.0, $-$18.5, $-$19.0, $-$19.5, $-$20.0, $-$20.5, and $-$21.0. The group masses in the catalogs have been estimated from velocity dispersions using the virial theorem. Velocity dispersions are obtained from the line-of-sight velocities of all detected galaxies in groups. The method assumes that the mass distribution follows the NFW profile \citep{1997ApJ...490..493N}. Using a mock galaxy catalog from the Millenium Simulation data, \citet{2014MNRAS.441.1513O} found that the rms error in the group masses recovered by the method compared to the true masses is around 0.3 dex. The dynamical mass estimates in groups with few members may be unreliable. For groups richer than four members, the group masses in flux- and volume-limited samples agree quite well and are reliable \citep{2014A&A...566A...1T}. For these reasons, we use only groups with at least five members in the main flux limited catalog for our study. We divide these groups into different volume limited samples with magnitude limits M$_{r,\mathrm{lim}} = -18.0, -19.0, -20.0, -21.0$ based on the SDSS $r$-band absolute magnitudes of their central galaxies. The galaxy with the highest luminosity in its group is classified as the central galaxy. We do not find any bias in galaxy stellar masses and group masses related to distance in all the volume limited samples defined by this approach. Table~\ref{table:1} shows the properties of the different volume limited samples. 
\begin{table}
\caption{Properties of different volume limited samples of groups extracted from the SDSS DR10 group catalog by \citet{2014A&A...566A...1T}. }              
\label{table:1}      
\centering                                      
\begin{tabular}{c c c c c}          
\hline\hline                        
Sample & $M_{r,\mathrm{lim}}$\tablefootmark{a} & $D_\mathrm{lim}$\tablefootmark{b} & N(total)\tablefootmark{c} & N(nrich > 4)\tablefootmark{d}\\    
       &  (mag)      &  ($h^{-1}$Mpc) & & \\
\hline                                   
M18 & $-$18 & 144 & 8657  & 1531 \\      
M19 & $-$19 & 213 & 20131 & 3676 \\
M20 & $-$20 & 330 & 37011 & 7077 \\
M21 & $-$21 & 487 & 34290 & 7268 \\
\hline                                             
\end{tabular}
\tablefoot{
\tablefoottext{a}{$r$-band limiting absolute magnitudes of the samples.}
\tablefoottext{b}{Upper distance limit of the samples.}
\tablefoottext{c}{Total number of groups in the samples.}
\tablefoottext{d}{Number of groups in the samples with richness greater than 4.}
}
\end{table}  

\subsection{Galaxy properties}
Stellar masses, star formation rates, and specific star formation rates of galaxies in groups used in this work come from measurements made by the MPA-JHU team on SDSS DR8 spectra available at \url{https://www.sdss3.org/dr10/spectro/galaxy_mpajhu.php}. They calculated stellar masses by fitting the observed optical ($u$, $g$, $r$, $i$, and $z$) spectral energy distributions to a large grid of models from \citet{2003MNRAS.344.1000B} spanning a large range in star formation histories. They also provide a likelihood distribution obtained from $\chi^2$ for the stellar mass for each galaxy. We take the median of the distribution as the best estimates for stellar masses in galaxies.

The star formation rates (SFRs) within the galaxy fiber aperture were computed using the nebular emission lines as described in \citet{2004MNRAS.351.1151B} and that outside of the fiber by using the galaxy photometry following \citet{2007ApJS..173..267S}. For AGN and galaxies with weak emission lines, the SFRs were estimated from the photometry. The specific SFR (SFR divided by the stellar mass) were computed by combining the SFR and stellar mass likelihood distributions as outlined in Appendix A of \citet{2004MNRAS.351.1151B}. They provide both the fiber and the total SFR and specific SFR at the median and 2.5$\%$, 16$\%$, 84$\%$ and 97.5$\%$ of the probability distribution function. We take the median as the best estimates for total SFR and specific SFR.

The morphological classification of galaxies into spirals and ellipticals is based on various criteria related to optical light profiles, optical colors, and probability \citep[][]{2011A&A...529A..53T}. The main classification is based on the values of de Vaucouleurs component ($f_{deV}$) in the best-fit composite model and the exponential fit axis ratio ($q_{exp}$) obtained for the light profiles of galaxies in the SDSS $r$-band. These values can be obtained from the SDSS public database. All galaxies, where $q_{exp} < 0.4$ or $q_{exp} < 0.9 - 0.8 f_{deV}$, are classified as spirals. The remaining galaxies are then classified into spirals or ellipticals by using SDSS $u$-$r$ and $g$-$r$ colors \citep[see][for details]{2011A&A...529A..53T}. The residual galaxies, where $f_{deV} > 0.7$ and $q_{exp} > 0.7$, are classified as ellipticals. In our analysis, only those spiral or elliptical galaxies classified using \citet[][]{2011A&A...529A..53T} scheme are taken for which the probability of being an early or late type galaxy as estimated in  \citet{2011A&A...525A.157H} is at least 50$\%$.

\section{Methods}
\label{sec:methods}
\subsection{Defining large-scale galaxy environments}
Several methods like minimal spanning tree algorithm \citep{1985MNRAS.216...17B,2014MNRAS.438..177A}, subspace constrained mean shift algorithm, and tidal tensor prescription \citep{2007MNRAS.375..489H,2015MNRAS.448.3665E} have been used to find filamentary structures in galaxy distribution. In this paper, we study galaxy filaments detected by marked point processes with interactions, also called Bisous model \citep[][and references therein]{2016A&C....16...17T}. The method is based on the assumption that galaxies may be assembled randomly into small cylinders which may merge with other neighboring cylinders having similar orientations to form filaments \citep{2014MNRAS.438.3465T}. Unlike other methods, this approach also provides the probability of the detected structures using the Bayesian framework. The detected filaments are found to follow the underlying velocity field in both simulations \citep{2014MNRAS.437L..11T} and observations \citep{2015MNRAS.453L.108L}. Using this method, \citet{2014MNRAS.438.3465T} constructed the cosmic web filament catalog from the SDSS DR10 data assuming the characteristic radius of the filaments as 0.5~$h^{-1}$Mpc. The galaxy catalog being flux-limited, this scale was chosen for reliability of the detected filaments over the whole distance range up to the redshift 0.2. We crossmatch the SDSS DR10 galaxy catalog with cosmic web filament catalog to find the galaxies that lie inside and outside of filaments. Galaxies are defined to be within the filamentary structures if the distances of galaxies from the axes of the filamentary cylinders are less than 0.5~$h^{-1}$Mpc. Using this criterion, about 20\% of galaxies in the main SDSS DR10 galaxy catalog are found within filaments\footnote{Fraction of galaxies in filaments depends on the success of filament finding algorithm. In nearby Universe ($z \leq$ 0.1), where the filaments are more complete, the fraction of galaxies in filaments is around 50$\%$ \citep[see][]{2014MNRAS.438.3465T}.}. \citet{2016A&C....16...17T} have found that most of the galaxies are located close to filaments and are usually closer than 0.4 $h^{-1}$Mpc from the filament spines. We also find that our results in this paper remain unchanged when the filament radius is increased to 1$h^{-1}$Mpc. Also, the increase of filament radius to 1 $h^{-1}$Mpc does not significantly affect the differences between galaxy and group properties in and outside of filaments.    

We also use the SDSS $r$-band luminosity density field of galaxies constructed by applying 8~$h^{-1}$Mpc smoothing kernel and 1~Mpc grid size to characterize the large-scale environments of galaxies and groups. This method is particularly efficient in finding voids and superclusters in the cosmic web \citep{2007A&A...462..811E,2012A&A...539A..80L,2012A&A...545A.104L}. We divide the groups in volume limited samples into three different large-scale environments based on their $r$-band 8~$h^{-1}$Mpc smoothed luminosity density values (Den8). The D1 sample contains groups in underdense void like regions having density values less than 1.5 \citep{2009A&A...495...37T,2016A&A...590A..29P}. The D2 sample consists of groups with density values in the range 1.5$<$Den8$<$5. The D3 sample is populated by groups in overdense supercluster like regions with density values greater than 5 \citep{2007A&A...462..811E,2012A&A...539A..80L}.

\subsection{Estimating error limits}
The error limits on any measured values in this study are estimated by using the bootstrap resampling technique \citep{efron1979,EFRON01121981,trove.nla.gov.au/work/26411468}. Using this technique, we draw galaxies or groups randomly from the main sample with repetitions and measure the mean properties. We repeat this process 1000 times, i.e, we have 1000 different estimates of the mean values of a measured parameter. We then calculate the mean of the distribution and take the error limit as $\pm$1$\sigma$ of the mean.  

\subsection{Group mass functions}
\label{sec:mass functions}
Groups in the volume limited samples are divided into logarithmic group mass bins to estimate the group mass functions. The total comoving volumes of the different volume limited samples with absolute magnitude limits $-$18.0 (M18), $-$19.0 (M19), $-$20.0 (M20), $-$21.0 (M21) are 1.517 $\times$ 10$^6$, 6.541 $\times$ 10$^6$, 24.27 $\times$ 10$^6$, and 83.78 $\times$ 10$^6$\,$h^{-3}$\,Mpc$^{3}$, respectively \citep{2013MNRAS.436..380N}. The group mass function is given by

\begin{equation}
\mathrm{\phi(\log \mathit{M_{g}}) d(\log \mathit{M_{g}})} =  c \times \mathrm{\frac{\sum_\mathit{i} I_{\it(M_{g},M_{g}+dM_{g})} (\log \it M_{{g},i})}{\mathit{V}}} 
.\end{equation}\\
Here, $c$ is the ratio of the number of groups in the whole catalogue to the number of groups used for mass function estimates. I$_A(x)$ is the indicator function that selects the groups belonging to a particular group mass bin (taken as $10^{0.5}$ M$_\odot$) and the sum runs over all groups in the sample. $V$ is the total comoving volume of the volume limited sample which is used to construct the mass function. We calculate the error limits by using the bootstrap resampling technique. We calculate the number densities at each group mass bin for 1000 random bootstrap samples. Then, we calculate the mean densities and 1$\sigma$ errors of the mean for all the mass bins to get the error limits.

\begin{figure*}[htbp]
   \centering
   \includegraphics[width=\hsize]{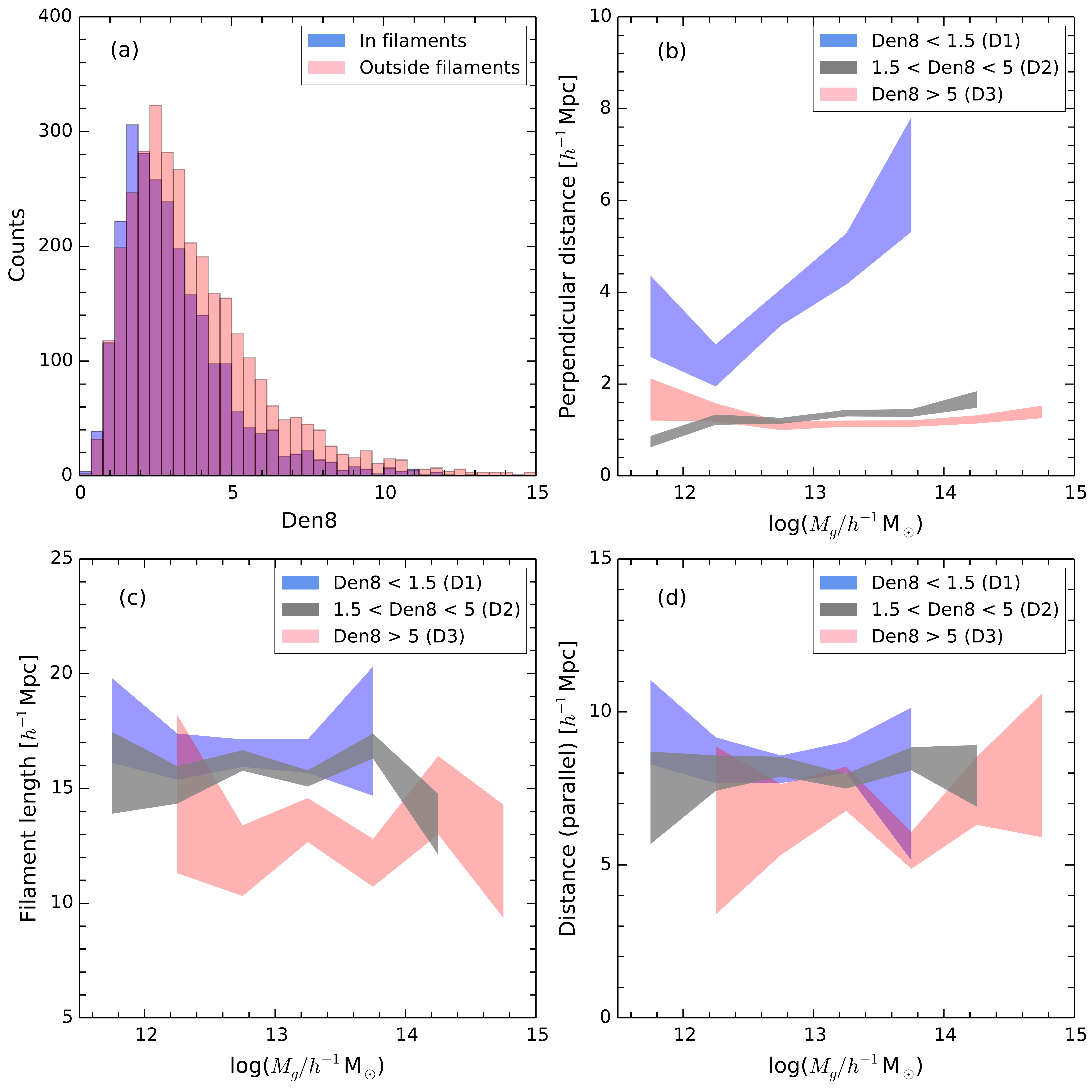}
      \caption{(a) 8~$h^{-1}$Mpc smoothed luminosity density distribution of groups inside (blue) and outside (light red) of filaments in M20 volume limited sample. Purple color represents overlapping distribution. (b) Mean perpendicular distances of central galaxies outside of filaments measured from the filament axes plotted as a function of group masses in different environments defined by the luminosity density field. (c) Mean length of filaments as a function of group masses in different environments within filaments. (d) Mean distances of central galaxies along the filament axes as a function of group masses in different environments defined by the luminosity density field. 
              }
         \label{Den8 distribution: Volume limited}
   \end{figure*}   
       \begin{figure}[htbp]
   \centering
   \includegraphics[width=\hsize]{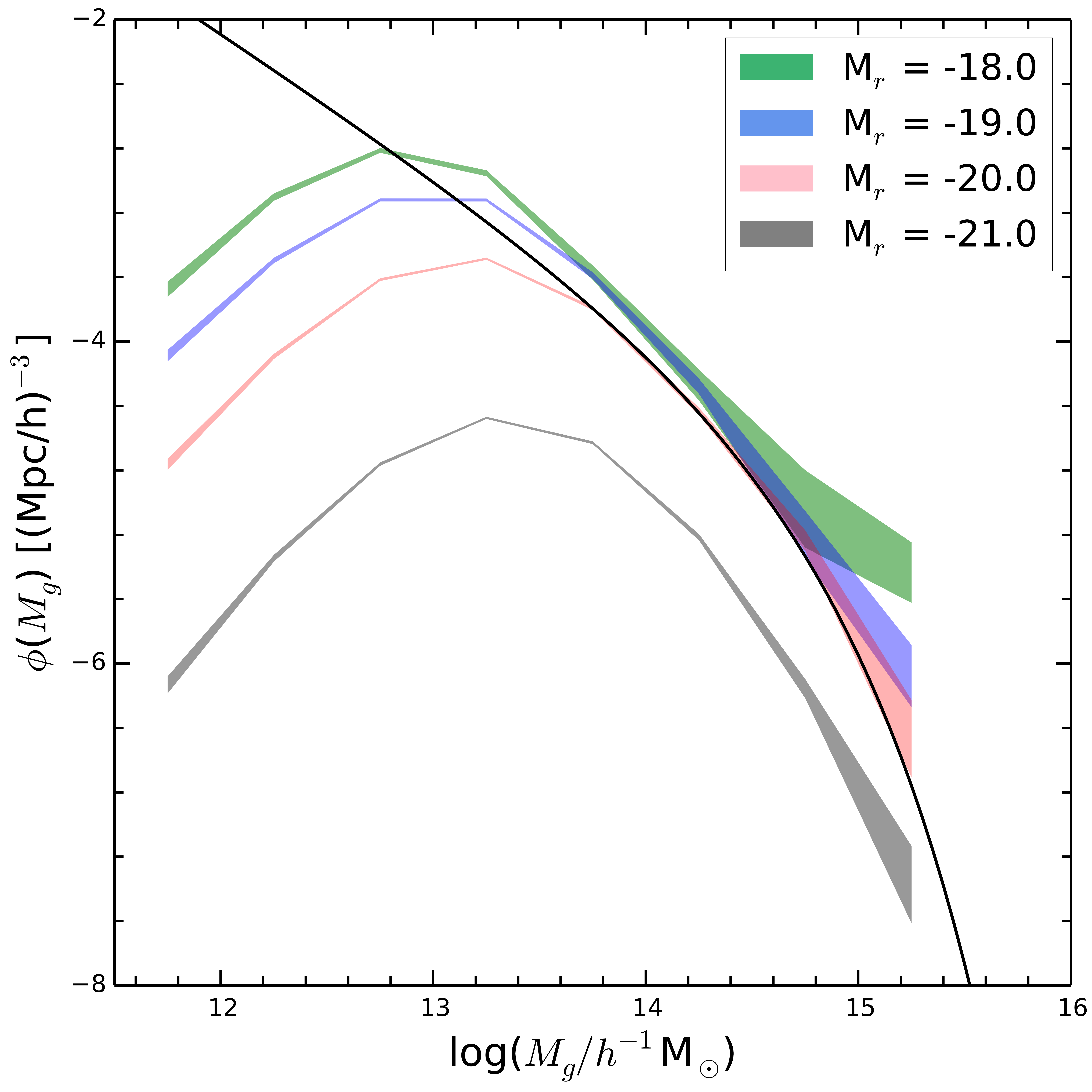}
      \caption{The total group mass functions in different volume limited group samples. The error limits are estimated using the bootstrap resampling technique. The mass functions are incomplete at the low mass end. The solid line represents the halo mass function from N-body simulations by \citet{2001MNRAS.323....1S}. 
              }
         \label{Total group mass functions: Volume limited}
   \end{figure}

\section{Group properties in different environments}
\label{sec:results(groups)}
\subsection{Groups in filaments}
Groups span a wide range of luminosity densities with intermediate densities being the most abundant and the distributions tail off towards underdense and overdense regions (Fig.~\ref{Den8 distribution: Volume limited}a). Groups in filaments tend to have lower luminosity densities than those outside of filaments but still have a broad range of luminosity density values. Filamentary structures are heterogenous environment in terms of luminosity densities. The environment of galaxies within filaments may span a wide range of luminosity densities characteristic of voids and superclusters. In the Bisous model, the detection of filaments is not directly related to the environmental density, rather the Bisous model is looking for structures where galaxies are arranged in filamentary configurations. Hence, the detection of filaments in superclusters and voids is expected behavior in the Bisous model. The large-scale environment based on luminosity density should be fixed in order to study the unbiased effect of filamentary structures on the properties of galaxies and groups. Groups are assumed to be in filament regions if their central galaxies lie within the filaments. 

Groups in very low-density environments outside of filaments are far away from the filament axes and the mean perpendicular distance increases with group mass (Fig.~\ref{Den8 distribution: Volume limited}b). However, groups in high-density environments lie on an average closer than 2~$h^{-1}$Mpc from the filament axes. Groups in high-density environments tend to reside in shorter filaments than those in low-density environments (Fig.~\ref{Den8 distribution: Volume limited}c). Along the axes of the filaments, there is no preferential arrangement of groups in terms of mass i.e., groups have a random distribution of masses along the filament axes (Fig.~\ref{Den8 distribution: Volume limited}d).
 
\subsection{Group mass functions}
\label{sec:results(mass functions)}

Figure \ref{Total group mass functions: Volume limited} shows the total group mass functions obtained in different volume limited samples and comparison with the halo mass function from N-body simulations by \citet{2001MNRAS.323....1S}. The group mass functions decrease sharply at the low mass end ($M_g < 13~h^{-1}$M$_\odot$) for all volume limited samples compared to the halo mass function by \citet{2001MNRAS.323....1S}. This is due to the resolution limit of the survey and our selection criterion of groups where we remove poor groups with richness less than five members, which generally have lower masses. The amplitudes of the mass functions towards the low mass end gradually decrease as we go towards the brighter magnitude limits. This is because brighter central galaxies tend to reside in more massive groups. The group mass functions in different volume limited samples agree well towards the massive end for M18, M19, and M20 samples. The M21 sample has lower normalization at the massive end which is due to the incompleteness of the sample. The group mass function from the M20 sample agrees quite well with the simulation result towards the massive end. For better statistical analysis on the dependence of galaxy and group properties on the large-scale environment, we use the M20 sample as it contains higher number of groups than M18 and M19 samples and is more complete than the M21 sample. 

\begin{figure*}[htbp]
   \centering
   \includegraphics[width=\hsize]{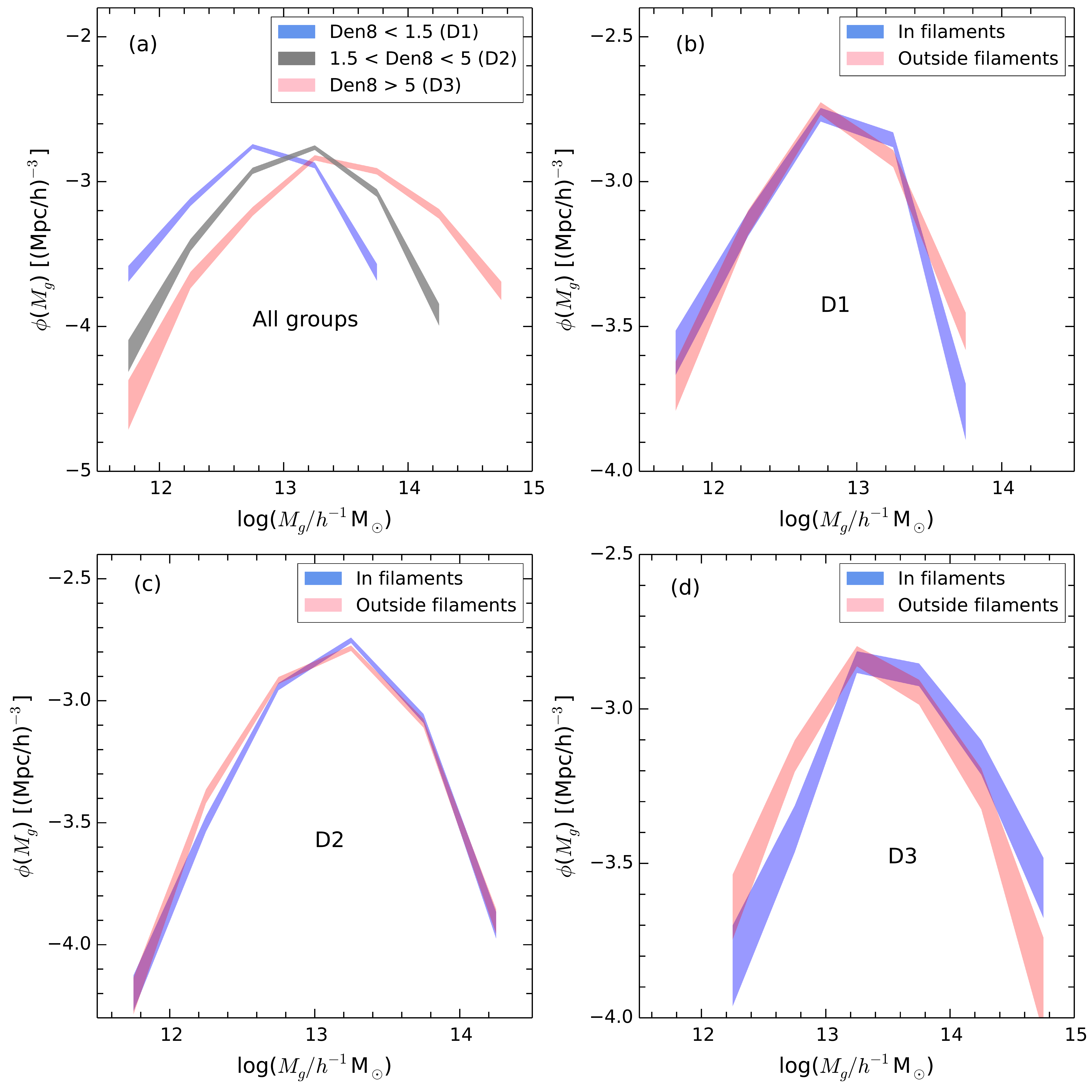}
      \caption{(a) Mass function of groups in different large scale environments defined by the luminosity density field. (b) Mass function of groups inside and outside filaments with luminosity density values Den8 < 1.5. (c) Mass function of groups inside and outside filaments with luminosity density values 1.5 < Den8 < 5. (d) Mass function of groups inside and outside filaments with luminosity density values Den8 > 5. 
              }
         \label{Group mass functions: Den8}
   \end{figure*}

We also construct the group mass functions in different large-scale environments defined by the luminosity density field (Fig.~\ref{Group mass functions: Den8}a). For direct comparison, the group mass functions are constructed by taking randomly equal number of groups from all the samples. The error limits are calculated in similar way as explained in Section~\ref{sec:mass functions}. The group mass functions in different environments based on the luminosity density field are different in all volume-limited samples. The turnover mass is the smallest in D1 and increases as we go towards higher densities D2 and D3. Our result that groups in high-density large scale environments are more massive also agrees with similar hydrodynamical simulation studies in \citet{2009MNRAS.399.1773C} although we use a larger environmental scale. 

To study the effect of the cosmic web filamentary structures on the group mass functions, we compare the mass functions of groups inside and outside of filaments for three different luminosity density ranges: D1 (Fig.~\ref{Group mass functions: Den8}b), D2 (Fig.~\ref{Group mass functions: Den8}c), and D3 (Fig.~\ref{Group mass functions: Den8}d). The mean densities and error limits are calculated in the same way as explained in Section~\ref{sec:mass functions}. The mass functions are shown for only those bins which contain at least ten groups. The mass functions of groups show slight enhancement at lower masses ($M_g < 10^{13} h^{-1}$ M$_\odot$) outside of filaments compared to that in filaments at high densities (Fig.~\ref{Group mass functions: Den8}d). Similar enhancement in mass functions is also seen outside of filaments at intermediate densities in the mass range $10^{12} < M_g < 10^{12.5} h^{-1}$ M$_\odot$ (Fig.~\ref{Group mass functions: Den8}c). At the high mass end, the mass function also shows an enhancement for groups at the lowest densities outside of filaments compared to that in filaments in the mass range, $M_g > 10^{13.5} h^{-1}$ M$_\odot$ (Fig.~\ref{Group mass functions: Den8}b). In contrast, at the highest densities, the mass function of groups tend to have higher amplitude inside filaments compared to that outside of filaments at group masses above $10^{13.5} h^{-1}$ M$_\odot$ (Fig.~\ref{Group mass functions: Den8}d).
      
    \begin{figure*}[htbp]
     \centering
  \includegraphics[width=\hsize]{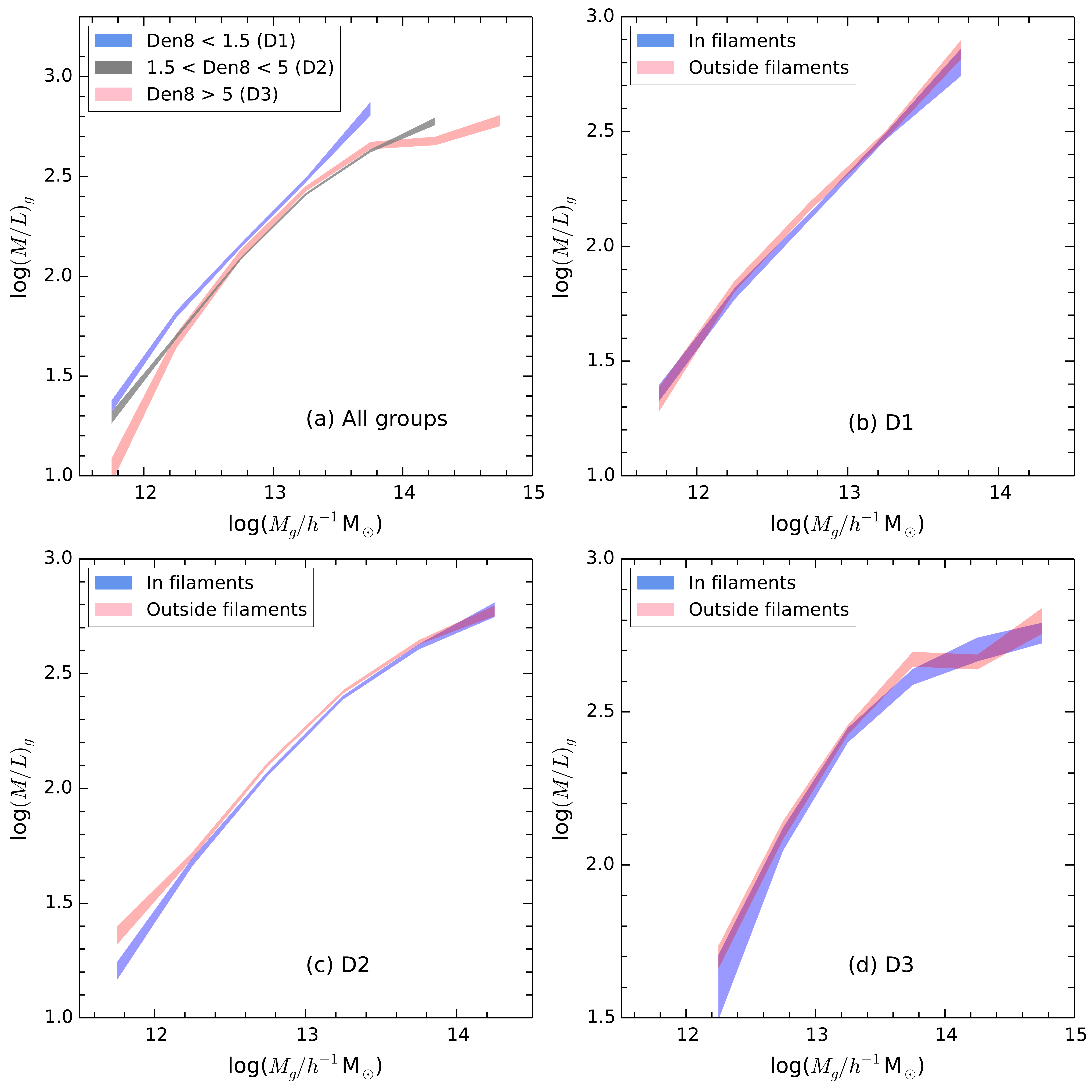}
      \caption{(a) Mass-to-light-ratio of groups as a function of group mass in different environments defined by the luminosity density field. (b) Mass-to-light-ratio of groups inside and outside filaments with luminosity density values Den8 < 1.5. (c) Mass-to-light-ratio of groups inside and outside filaments with luminosity density values 1.5 < Den8 < 5. (d) Mass-to-light-ratio of groups inside and outside filaments with luminosity density values Den8 > 5.
              }
         \label{Group mass to light ratio: Den8}
   \end{figure*}
\subsection{Group mass to light ratio}

For the group total luminosities, we used the luminosity values from  the original flux-limited catalog where the group luminosities have been corrected to take into account the missing galaxies that lie outside the flux limit of the survey \citep{2014A&A...566A...1T}. The mass-to-light $(M/L)_g$ ratio of a group is obtained by dividing the total mass of the group with its total luminosity in the SDSS $r$ band. The mass-to-light ratio increases with the group mass in all environments (Fig.~\ref{Group mass to light ratio: Den8}a). Using cross-correlation weak lensing, \citet{2009ApJ...703.2232S} also found that the $(M/L)_g$ ratio within $r_{200}$, scales with cluster mass as a power law with index 0.33$\pm$0.02. The $(M/L)_g$ ratio curve tends to flatten at higher masses in high-density environments, whereas it continues rising in low-density environments, i.e. the power law indices vary between high- and low-density environments. At fixed group mass, groups in low-density environments tend to have higher mass to light ratios that those in low-density environments, i.e., groups in high-density environments are brighter than those in low density environments . The difference is higher between D1 and D2 or D3 sample. Between D2 and D3 samples, the difference between group mass to light ratio is not so clear.

In order to study the influence of the cosmic web filaments on the $(M/L)_g$ ratio in groups, we compare the $(M/L)_g$ ratio of groups inside and outside of filaments with similar luminosity densities and group masses. We plot the $(M/L)_g$ ratio of groups as a function of group masses for three different luminosity density ranges: D1 (Fig.~\ref{Group mass to light ratio: Den8}b), D2 (Fig.~\ref{Group mass to light ratio: Den8}c), and D3 (Fig.~\ref{Group mass to light ratio: Den8}d). Error limits are calculated by using the bootstrap resampling technique. The $(M/L)_g$ ratio of groups are similar both inside and outside of filaments in D1 and D3 samples. In the D2 sample, the $(M/L)_g$ ratio of groups outside of filaments is slightly higher than those within filaments especially at lower masses. This implies that, at intermediate luminosity density range, similar mass groups in filaments are more luminous than those outside of filaments. 
        \begin{figure*}[htbp]
  \centering
   \includegraphics[width=\hsize]{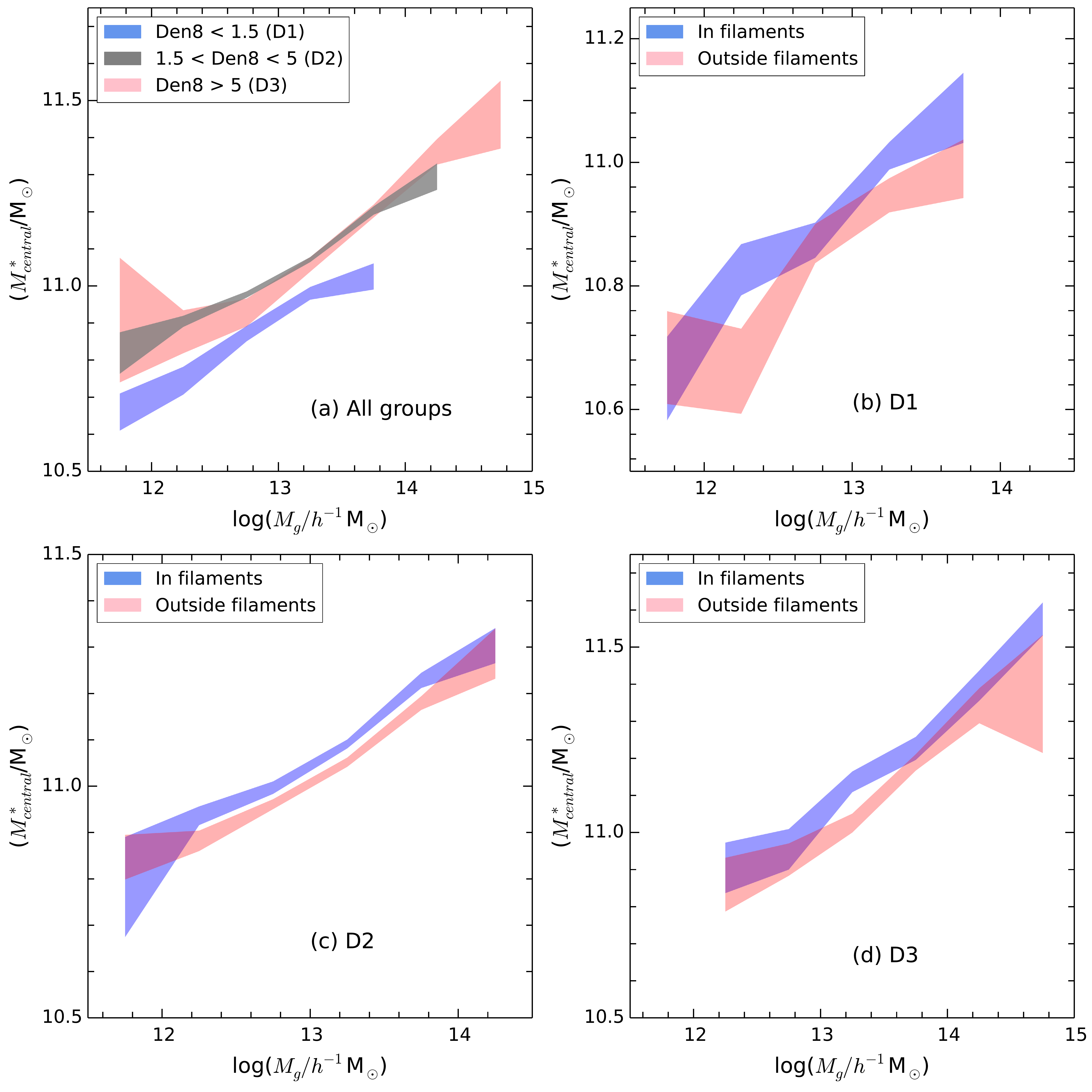}
      \caption{(a) Central galaxy stellar mass as a function of group mass in different large-scale environments defined by the luminosity density field. The error limits are estimated using the bootstrap resampling technique. (b) Central galaxy stellar mass as a function of group mass for groups in and outside of filaments having luminosity densities, Den8 < 1.5. (c) Central galaxy stellar mass as a function of group mass for groups in and outside of filaments having luminosity densities, 1.5 < Den8 < 5. (d) Central galaxy stellar mass as a function of group mass for groups in and outside of filaments having luminosity densities, Den8 > 5. 
              }
         \label{Central galaxy: stellar mass}
   \end{figure*}    

   \begin{figure*}[htbp]
  \centering
   \includegraphics[width=\hsize]{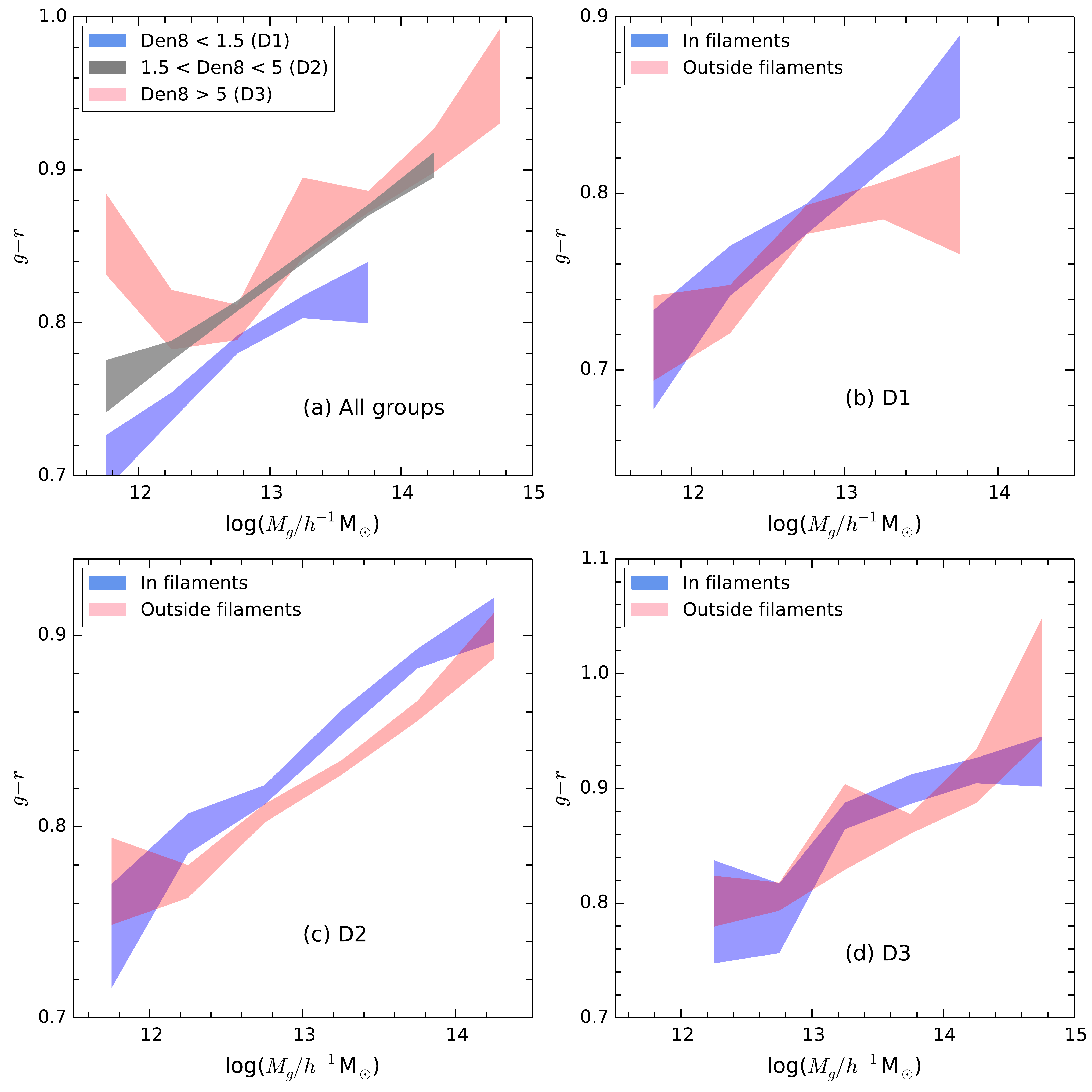}
      \caption{(a) $g$-$r$ color of central galaxies as a function of group mass in different large-scale environments. (b) $g$-$r$ color of central galaxies as a function of group mass for groups in and outside of filaments with luminosity densities, Den8 < 1.5. (c) $g$-$r$ color of central galaxies as a function of group mass for groups in and outside of filaments with luminosity densities, 1.5 < Den8 < 5. (d) $g$-$r$ color of central galaxies as a function of group mass for groups in and outside of filaments with luminosity densities, Den8 > 5.
              }
          \label{central galaxy: g-r color}
   \end{figure*}

\section{Properties of central galaxies}
\label{sec:results(central galaxies)}
In this section, we study the effect of the large-scale environment on the stellar mass, color, and specific star formation rate of central galaxies. Central galaxies are those with the highest luminosities in their respective groups.

\subsection{Stellar mass}
The stellar mass of the central galaxy is an increasing function of group mass in different environments defined by the luminosity density field. At a fixed group mass, groups in high-density environments have higher central galaxy stellar masses than those in low density environments (Fig.~\ref{Central galaxy: stellar mass}a). The lowest density sample (D1) differs the most from the others. At fixed group masses, higher large-scale density (D2 and D3) samples show no differences in stellar masses between them across the whole group mass range. 
 
To quantify the effect of cosmic web filaments on the central galaxy stellar mass, we plot the stellar mass of the central galaxy as a function of the group mass for groups inside and outside of filaments in three different luminosity density ranges: D1 (Fig.~\ref{Central galaxy: stellar mass}b); D2 (Fig.~\ref{Central galaxy: stellar mass}c); D3 (Fig.~\ref{Central galaxy: stellar mass}d). In the D3 sample, the stellar mass of central galaxies in filaments are clearly higher than those outside of filaments in the mass range 10$^{13-13.5}$ $h^{-1}$ M$_\odot$. At higher masses, there is only a small overlap and the mean central stellar masses (without errors) are always higher in filaments. At lower masses, there is a considerable overlap. In the D2 sample, we find that the stellar mass of central galaxies within filaments have higher stellar masses than those outside of filaments in the mass range 10$^{12-14}$ $h^{-1}$ M$_\odot$ . At lower and higher group masses, stellar masses of the central galaxies are similar inside and outside of filaments. In D3 sample, the stellar mass of central galaxies are similar in the mass ranges $M_g <$ 10$^{12}$ $h^{-1}$ M$_\odot$ and 10$^{12.5-13}$ $h^{-1}$ M$_\odot$. At other group masses, stellar masses of the central galaxies in filaments are higher than those outside of filaments. In general, at a fixed group mass, central galaxies with similar luminosity density values in filaments have higher stellar masses than those outside of filaments.
  
\subsection{g-r color}
Central galaxy color tends to follow similar trend as the stellar mass in different large-scale environments defined by luminosity densities. In general, central galaxies get redder on increasing group masses in all environments (Fig.~\ref{central galaxy: g-r color}a). At a fixed group mass, central galaxies in low-density environments (D1 sample) are bluer than those in other high-density environments in the whole mass range. Denser environments with intermediate (D2 sample) and very high (D3 sample) luminosity densities, show no clear differences in central galaxy colors. However, at very low masses ($M_g < 12 h^{-1} M_\odot$), central galaxies in the D3 sample also have redder colors than those in the D2 sample. The results agree with \citet{2006ApJ...638L..55Y} and \citet{2008ApJ...687..919W}, where they showed that groups with the same mass are less clustered if their central galaxies are bluer. If we associate central galaxy color to age, this result agree with \citet{2014MNRAS.443.3107L} where they find that old central galaxies have a higher clustering amplitude at scales $> 1~h^{-1}$Mpc than young central galaxies of equal host halo mass. This may be interpreted as an assembly-type bias found earlier in semi-analytical models \citep{2007MNRAS.374.1303C,2011MNRAS.412.1283L}.  

In order to study the impact of cosmic web filaments on $g-r$ color of central galaxies, we plot the mean color of the central galaxy as a function of group mass for groups inside and outside of filaments in three different luminosity density ranges: D1 (Fig.~\ref{central galaxy: g-r color}b); D2 (Fig.~\ref{central galaxy: g-r color}c); D3 (Fig.~\ref{central galaxy: g-r color}d). We find differences in $g-r$ color of central galaxies inside and outside of filaments with similar luminosity densities. The D3 sample shows no differences in central galaxy colors inside and outside of filaments. In the D2 sample, at a fixed group mass, central galaxies in filaments are redder than those outside of filaments in the mass range 10$^{12-14} h^{-1} M_\odot$. At very low and very high group masses, central galaxies in filaments have similar $g-r$ colors compared to those outside filaments. In the D1 sample, at a fixed group mass, central galaxies in filaments have redder colors than those outside of filaments in the mass range 10$^{13-14} h^{-1} M_\odot$. At lower masses, central galaxies have similar $g-r$ colors irrespective of their location. The K-S test shows very low probabilities that the mean $g-r$ color of central galaxies is similar inside and outside of filaments at fixed large-scale environment (Table.~\ref{table:2}) and group mass (Table.~\ref{table:3}). This shows the reliability of the observed differences.

\subsection{Specific star formation rates}
\begin{figure*}[htbp]
  \centering
   \includegraphics[width=\hsize]{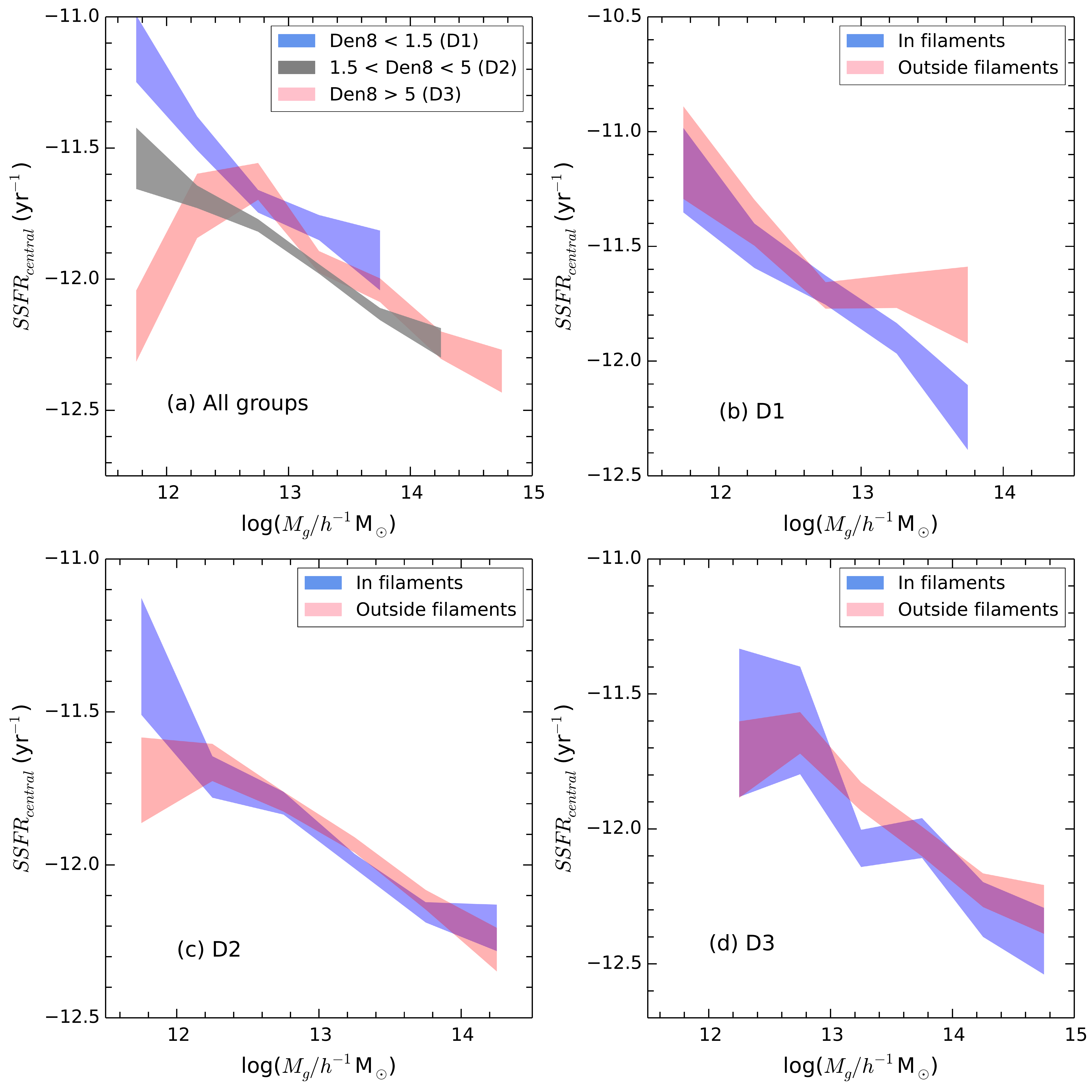}
      \caption{(a) Specific star formation rate of central galaxies as a function of group mass in different large-scale environments. (b) Specific star formation rate of central galaxies as a function of group mass for groups in and outside of filaments with luminosity densities, Den8 < 1.5. (c) Specific star formation rate of central galaxies as a function of group mass for groups in and outside of filaments with luminosity densities, 1.5 < Den8 < 5.(d) Specific star formation rate central galaxies as a function of group mass for groups in and outside of filaments with luminosity densities, Den8 > 5.
              }
         \label{central galaxy: ssfr}
   \end{figure*}

In general, the specific star formation rates of the central galaxies decrease as a function of group mass in all environments defined by the luminosity density field. At a fixed group mass, central galaxies in low-density environments have higher specific star formation rates compared to high density environments for groups with masses lower than 10$^{12.5} h^{-1} M_\odot$ or higher than 10$^{13} h^{-1} M_\odot$ (Fig.~\ref{central galaxy: ssfr}a). The SSFR of central galaxies for D1 and D3 samples are comparable in the mass range 10$^{12.5-13} h^{-1} M_\odot$ but higher than that for the D2 sample. The difference is better seen between D1 and D2 samples. At very low masses, the SSFRs of central galaxies are clearly different in all environments, with galaxies in the lowest density range having the highest SSFRs and galaxies in the highest density range having the lowest SSFRs. 
      \begin{figure*}[htbp]
  \centering
   \includegraphics[width=\hsize]{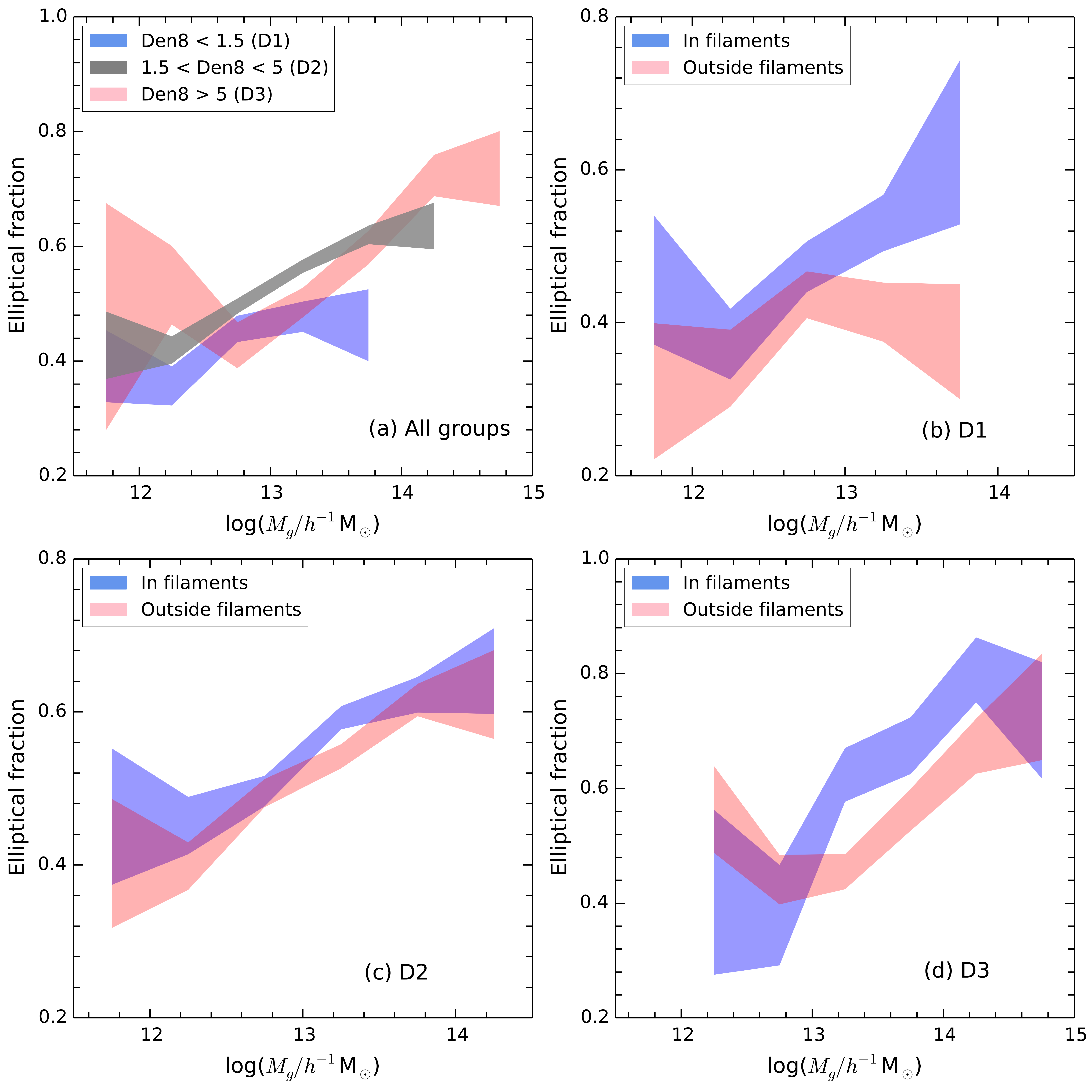}
      \caption{(a) Elliptical fraction of central galaxies as a function of group mass in different large-scale environments. (b) Elliptical fraction of central galaxies as a function of group mass for groups in and outside of filaments with luminosity densities, Den8 < 1.5. (c) Elliptical fraction of central galaxies as a function of group mass for groups in and outside of filaments with luminosity densities, 1.5 < Den8 < 5.(d) Elliptical fraction of central galaxies as a function of group mass for groups in and outside of filaments with luminosity densities, Den8 > 5.
              }
         \label{central galaxy: elliptical fraction}
   \end{figure*}    
      \begin{figure*}[htbp]
  \centering
   \includegraphics[width=\hsize]{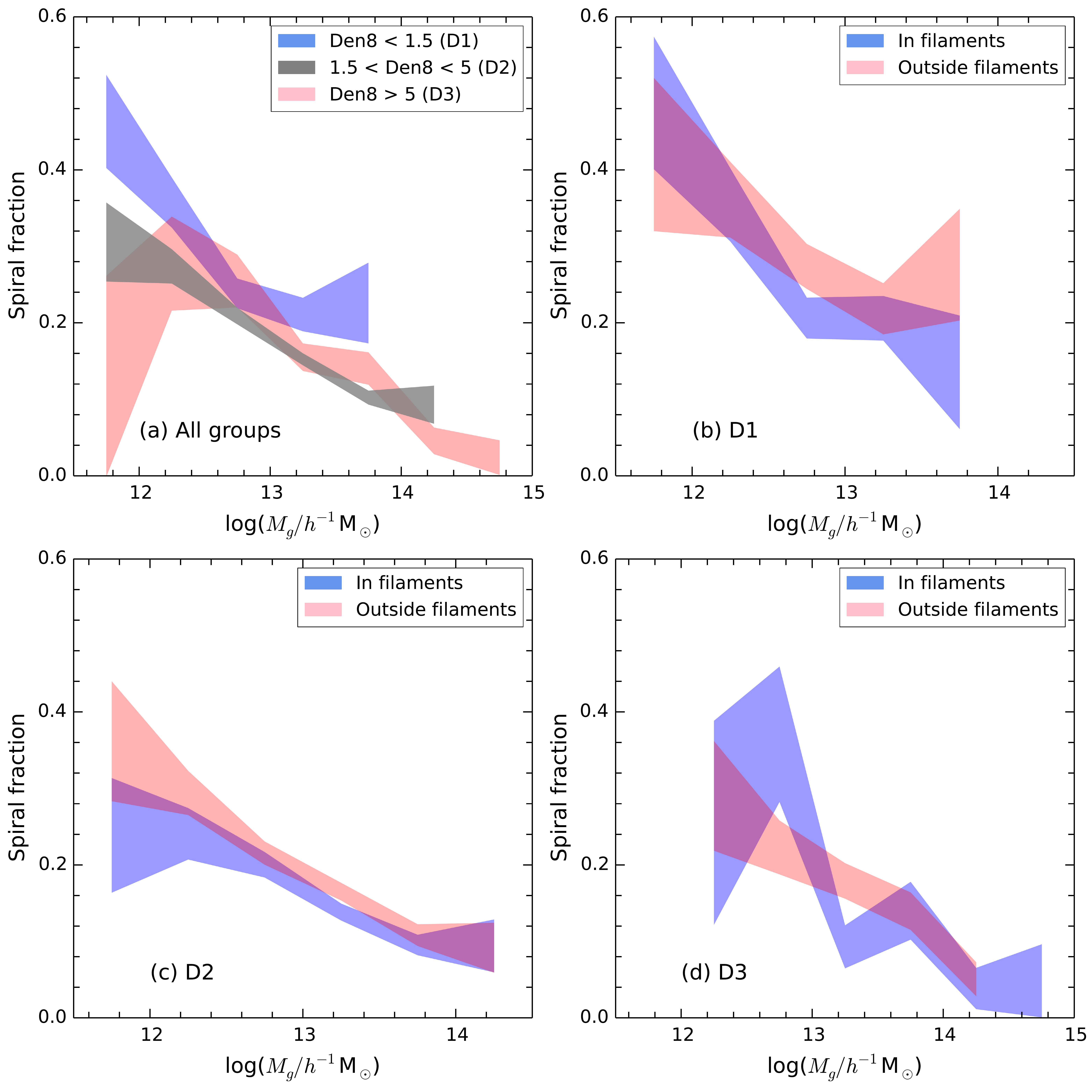}
      \caption{(a) Spiral fraction of central galaxies as a function of group mass in different large-scale environments. (b) Spiral fraction of central galaxies as a function of group mass for groups in and outside of filaments with luminosity densities, Den8 < 1.5. (c) Spiral fraction of central galaxies as a function of group mass for groups in and outside of filaments with luminosity densities, 1.5 < Den8 < 5.(d) Spiral fraction of central galaxies as a function of group mass for groups in and outside of filaments with luminosity densities, Den8 > 5.
              }
         \label{central galaxy: spiral fraction}
   \end{figure*}  
The mean SSFR of the central galaxy is also plotted as a function of group mass for groups inside and outside of filaments in three different luminosity density ranges: D1 (Fig.~\ref{central galaxy: ssfr}b); D2 (Fig.~\ref{central galaxy: ssfr}c); D3 (Fig.~\ref{central galaxy: ssfr}d). In D3 sample, we do not find differences in the SSFRs of central galaxies inside and outside of filaments. In the D2 sample, at a fixed group mass, central galaxies in filaments have slightly higher SSFRs than those outside of filaments in the mass range $M_g <$ 10$^{12} h^{-1} M_\odot$. At higher masses, central galaxies have similar SSFRs inside and outside of filaments. In D1 sample, at a fixed group mass, central galaxies in filaments have lower SSFRs than those outside of filaments in the mass range $M_g >$ 10$^{13} h^{-1} M_\odot$. At lower masses, the SSFRs of central galaxies are similar inside and outside of filaments. The K-S test shows very low probabilities that the mean SSFR of central galaxies are similar inside and outside of filaments at fixed large-scale environment (Table.~\ref{table:2}) and group mass (Table.~\ref{table:3}). This shows that our results are reliable.

\subsection{Morphology}
\label{sec:results(morphology)}

The fraction of central galaxies with elliptical morphology increases with group masses in all large-scale environments defined by the luminosity density field (Fig.~\ref{central galaxy: elliptical fraction}a). In contrast, the spiral fraction decreases with group masses (Fig.~\ref{central galaxy: spiral fraction}a). Similar morphological trends of central galaxies were also found with halo mass by \citep{2012ApJ...746..160W} but their study does not consider large-scale environment. At a fixed group mass, groups in the high density sample (D2) have a higher fraction of elliptical galaxies than the low-density sample (D1). D1 and D3 samples have comparable elliptical fractions in the mass range 10$^{12.5-13.5} h^{-1} M_\odot$. D1 and D2 samples have comparable elliptical fraction in the mass range $M_g <$ 10$^{12} h^{-1} M_\odot$. The fraction of central galaxies with spiral morphology decreases with group masses in all large-scale environments defined by the luminosity density field (Fig.~\ref{central galaxy: spiral fraction}a). The spiral fraction in high-density environments (D2 or D3) is lower than that in low-density environments (D1) in most of the group mass bins. D1 and D3 samples have comparable spiral fraction in the mass range 10$^{12-13} h^{-1} M_\odot$.

We plot the mean morphological (elliptical or spiral) fraction of the central galaxy as a function of group mass for groups inside and outside of filaments in three different luminosity density ranges: D1, D2, and D3. In the D1 sample, at a fixed group mass, central galaxies in filaments have higher mean elliptical fraction than those outside of filaments in the mass range $M_g >$ 10$^{13} h^{-1} M_\odot$ (Fig.~\ref{central galaxy: elliptical fraction}b). In the D3 sample, the mean elliptical fraction is also clearly higher in the group mass range 10$^{13} h^{-1} M_\odot$<$M_g <$ 10$^{14.5} h^{-1} M_\odot$ (Fig.~\ref{central galaxy: elliptical fraction}d). In the D2 sample, the differences are unclear, but the mean elliptical fraction values without considering errors are always higher in filaments compared to those outside of filaments in the whole group mass range considered (Fig.~\ref{central galaxy: elliptical fraction}c). However, the differences in spiral fractions are unclear (Fig.~\ref{central galaxy: spiral fraction}). The K-S test shows very low probabilities that the elliptical fraction of central galaxies are similar inside and outside of filaments at fixed large-scale environment (Table.~\ref{table:2}) and group mass (Table.~\ref{table:3}). This shows the reliability of our results. Additionally, we also find that at fixed galaxy morphology, the differences in stellar mass, color, and SSFR of central galaxies inside and outside of filaments are not significant in groups with similar masses and large-scale environments.

\section{Discussion}
\label{sec:discussion}
\subsection{Halo mass assembly}
The mass distribution in the very early Universe was very homogeneous, but gravitational instabilities in the mass density field caused tiny overdensities to grow to form virialized dark matter halos. Small dark matter halos formed first, which grew either by accreting mass from their surroundings or by merging with other dark matter halos. \citet{2010ApJ...719..229G} studied the growth of dark matter halos using the Millennium and Millennium-II Simulations covering the ranges 10$^9$-10$^{15}$ M$_\odot$ in halo mass and 1-10$^5$ in merger mass ratio and found that all resolved mergers contribute about 60$\%$ of the total halo mass growth. They also found that, independent of halo mass, about 40$\%$ of the mass in halos comes from smooth accretion of dark matter that was never bound in smaller halos. The halo merger rates in high-density environments are found to be in higher than in low-density environments \citep{2009MNRAS.394.1825F}. This means that halos in high-density environments should have undergone more mergers during their evolution and hence should be more massive than halos formed at the same time in low-density environments. The comparisons of group mass functions in the Section~\ref{sec:results(mass functions)} follow this theoretically expected trend that high-density environments inhabit more massive halos than low-density environments. Unlike mergers and smooth accretion, \citet{2009MNRAS.398.1742H} predicted that the mass assembly of halos in the vicinity of massive halos may be slowed down by tidal effects due to the shear along the filaments feeding the massive halo. At the highest luminosity densities, filaments are found to have higher number densities of groups than the environment outside of filaments towards the high mass end of the group mass function (Fig.~\ref{Group mass functions: Den8}d). But, at the lowest luminosity densities, filaments are found to have lower number densities of groups than the environment outside of filaments towards the high mass end of the group mass functions (Fig.~\ref{Group mass functions: Den8}b). These results suggest that the tidal effects may influence the mass assembly of halos in filaments.

Simulation studies that have investigated the formation times of halos in different large-scale environments have found that halos in denser regions form earlier than in less dense regions \citep{2014ApJ...794...74J,2015ApJ...812..104T}. In our study, when comparing groups inside and outside of filaments, we fix the large-scale environment defined by their luminosity densities. Thus it is possible that the differences in the group mass functions are not simply related to the difference in formation times as assembly bias predicts. The observed results may be due to differences in mass assembly to halos in and outside of filaments driven by various physical processes like mergers, tidal effects and smooth accretion. 

\subsection{Group mass to light ratio}
Mass-to-light ratio increases systematically with an increasing spatial scale up to a few hundred kpc inside groups and clusters before approaching to a nearly constant value, the constancy extending up to tens of Mpc \citep{1995ApJ...447L..81B,2014MNRAS.439.2505B,2009ApJ...703.2232S}. This shows that light follows mass on large scales. Our results indicate that the relation between mass and light in virialized groups is not universal, but varies with large-scale environments defined by the luminosity density field (see Fig.~\ref{Group mass to light ratio: Den8}a). Groups in high-density environments contain more light per unit mass than those in low-density environments suggesting that intergalactic gas in groups in high-density environments condenses into stars more efficiently than in low-density environments. The differences may also be related to earlier halo formation times and higher abundances of filaments feeding galaxies in high-density environments \citep{2015ApJ...812..104T}.

Filaments in the cosmic web are enormous reservoirs of gas in the Universe most of which is cold with temperatures less than $10^5$ Kelvin \citep{2016MNRAS.455.2804S}. Hydrodynamic simulations suggest that relatively cool, unshocked gas flow along filaments of the cosmic web into dark-matter halos \citep{2005MNRAS.363....2K,2006MNRAS.368....2D,2014MNRAS.441.2923C} which provide fuel for star formation in galaxies. There are also few observational evidences that show that the cold gas flow from the cosmic web filaments can condense to form stars \citep{2015ApJ...810L..15S} and even galaxies \citep{2016ApJ...824L...5M}. In support to this scenario, groups in filaments are found to have more satellites than those outside of filaments \citep{2015ApJ...800..112G}. In our study, we find that mass-to-light ratio of groups inside filaments is lower than those outside of filaments with similar masses and large-scale densities i.e., groups in filaments are more luminous than outside of filaments (see Fig.~\ref{Group mass to light ratio: Den8}).The K-S test results show that there are very low probabilities that the distributions of mean mass to light ratio of groups inside and outside of filaments from bootstrapped samples are similar in each mass bins (see Table~\ref{table:2})  As we have fixed the dynamical mass, groups have similar halo environment and also by fixing the large-scale luminosity densities, the large-scale environment is also similar. The lower mass to light ratio of groups in filaments compared to those outside of filaments may therefore be due to higher galaxy content of groups in filaments. Also, galaxies of different morphological types may have different mass to light ratios \citep{1995ApJ...447L..81B}. The differences in galaxy morphological content in groups in and outside of filaments may also result in different mass to light ratio in groups with similar masses.                                                                                            

\subsection{Stellar mass assembly of central galaxies}
\label{mstar}
The stellar mass growth of galaxies within halos may occur either by accretion of gas on to the center of the gravitational potential wells of the dark matter or by merging with other galaxies. The build up of stellar mass in central galaxies may initially occur by accretion of gas and stars from its surroundings and in situ star formation, followed by galaxy mergers. In semi-analytical models, the galaxy merger rate is found to be higher in overdense regions \citep{2012ApJ...754...26J}. High-density environments also have more satellites \citep{2016A&A...590A..29P} that can feed the matter to central galaxies through mergers. In this paper, we find that central galaxies in high-density environments have higher stellar masses than those in low-density environments, even if the halo masses are similar (see Fig.~\ref{Central galaxy: stellar mass}a). 

We also find that at fixed group mass and large-scale environment, central galaxies in filaments have higher stellar mass than those outside of filaments (see Fig.~\ref{Central galaxy: stellar mass}b--d). The K-S test shows very low probabilities that the distributions of mean central galaxy stellar masses inside and outside of filaments from bootstrapped samples are similar in each mass bins (see Table~\ref{table:2}). However, if we also fix the central galaxy morphology, we find that the differences are insignificant. This means that the difference in stellar mass is related to central galaxy morphology. At fixed group mass and large-scale environment, a higher fraction of central galaxies in filaments are ellipticals. Elliptical galaxies are believed to form from mergers of spiral galaxies \citep{1977egsp.conf..401T}. This means that mergers may be more common with central galaxies in filaments compared to those outside of filaments. 

Hydrodynamical simulations predict that the accretion of gas from the cosmic web drives the growth of disk galaxies \citep{2006MNRAS.368....2D,2009Natur.457..451D,2012RAA....12..917S,
2014A&ARv..22...71S}. Also in simulations, the cold gas abundances in the cosmic web filaments are found to be higher than in voids and clusters \citep{2016MNRAS.455.2804S}. But, at fixed group mass and large-scale environment, we find that central galaxies in filaments with spiral morphology do not show clear enhancement in stellar mass and star formation activity compared to those outside of filaments with same morphology. This result indicates that the cold gas accretion from the cosmic web may not be the dominant mechanism for the stellar mass assembly of central galaxies. Also recent observations (like HALOGAS survey) investigating the cold gas accretion in nearby spiral galaxies have suggested qualitatively a very minor contribution of visible neutral hydrogen to the star formation fueling process \citep{2015IAUS..309...69H}.

\subsection{Morphology, color, and SSFR of central galaxies}
\label{color and ssfr}  
Galaxy morphology, color, and SSFR are correlated with each other, such that elliptical galaxies are generally redder, and have lower SSFR than spiral galaxies. It is also well known that the galaxy morphology \citep{1980ApJ...236..351D,1984ApJ...281...95P}, color \citep{2003ApJ...594..186B}, and star formation rate \citep{2002MNRAS.334..673L,2004MNRAS.348.1355B} depend on the local galaxy density, such that denser environments are populated by early-type, redder, and passive galaxies. We find that this is also true for large-scale environments defined by the luminosity density field (see Figs.~\ref{central galaxy: g-r color}a,~\ref{central galaxy: ssfr}a,~\ref{central galaxy: elliptical fraction}a, and ~\ref{central galaxy: spiral fraction}a). A possible explanation to these differences may be simply related to the effect of assembly bias such that galaxies in high-density environments are older than low-density environments. It is thus very a likely cause for stellar populations to be older and forming less stars in high-density environments.

We find that the morphology, color, and SSFR of central galaxies are also connected to filamentary structures in the cosmic web. The central galaxies in filaments are mostly ellipticals, redder, and more passive than those outside of filaments at fixed group mass and large-scale environment (see Figs.~\ref{central galaxy: g-r color},~\ref{central galaxy: ssfr},~\ref{central galaxy: elliptical fraction}). Comparing galaxies with similar luminosity densities (assuming that luminosity density reflects number density), we minimize the assembly bias effect and reduce possible differences due to the age of the central galaxy. The differences in colors and SSFRs disappear when comparing central galaxies with the same morphology inside and outside of filaments. This shows that the differences in mean color and SSFR seen inside and outside of filaments are mainly driven by higher abundances of elliptical galaxies in filaments and the morphological transformation of galaxies in filaments to elliptical types should be followed by turning off star formation. 

The reddening of color and low star formation activity in majority of elliptical galaxies compared to disk-type galaxies suggests that morphological transformation is often associated with change in color and star formation rate. Physical mechanisms like ram pressure stripping \citep{1972ApJ...176....1G}, strangulation \citep{1980ApJ...237..692L}, and harrassment \citep{1996Natur.379..613M} are believed to quench the star formation in galaxies but are efficient only in satellite galaxies. \citet{2014MNRAS.440..889S} found observational evidences that quenching in spiral galaxies occur when their external gas supply ends but continue forming stars from the residual gas. In elliptical galaxies, both the accreting gas and their reservoirs are consumed in a very short time scale. In central galaxies, mergers may induce burst of star formation and rapidly deplete cold gas via star formation and AGN feedback in a relatively short timescale making them red and passive \citep{2010ApJ...714L.108S}. Using high-resolution hydrodynamical simulation, \citet{2013MNRAS.433.3297D} have shown that AGN feedback can transform blue massive late-type central disc galaxies into red early-type galaxies. Recently, \citet{2016A&A...592A..30A} have also found that the optical AGN activity is related to the large-scale environment rather than the local environment. The AGN feedback is an efficient mechanism for quenching star-formation in massive galaxies. According to the model proposed by \citet{2016arXiv160707881A} based on simulations, non-linear interactions such as mergers in the cosmic web may separate the galaxy from its network of primordial filaments, cutting off the cold gas supply and shutting down star-formation in galaxies even without the presence of any black hole feedback. This mechanism quenches star-formation in galaxies at all mass ranges.     

\section{Conclusion}
\label{sec:conclusion}

We have used the SDSS DR10 groups, luminosity density field and cosmic web filaments to study the properties of groups and their central galaxies in different large-scale environments. The main results are summarized below:

\begin{enumerate}
\item Using the luminosity density field with 8~$h^{-1}$Mpc smoothing scale, we find that groups in high-density environments are more massive than those in low-density environments. At higher large-scale densities (Den8 > 5), the mass functions in filaments have higher amplitudes those outside of filaments at the massive end. At the lowest large-scale densities (Den8 < 1.5), the mass functions in filaments have lower amplitudes those outside of filaments at the massive end. 

\item At a fixed dynamical mass, groups in high-density environments are more luminous and their central galaxies have higher stellar mass than those in low-density environments with similar dynamical masses. On comparing groups inside and outside filaments in environments with similar luminosity densities and group masses, we find that groups in filaments are more luminous and their central galaxies have higher stellar mass than those outside of filaments. 

\item Using the luminosity density field with 8~$h^{-1}$Mpc smoothing scale, we find that central galaxies in groups in high-density environments have redder color and lower specific star formation rates than those in low-density environments in similar mass groups. At fixed halo and large-scale environment, we find that central galaxies in groups in filaments have redder color and lower specific star formation rates than those outside of filaments. 

\item Using the luminosity density field with 8~$h^{-1}$Mpc smoothing scale, we find that central galaxies in similar mass groups in high-density environments have higher abundances of elliptical galaxies than low-density environments. The abundances of spiral galaxies decrease with increasing large-scale densities. At fixed halo and large-scale environment, we find that central galaxies in filaments have higher elliptical fraction than outside of filaments. At fixed galaxy morphology, central galaxies in filaments show no clear differences in stellar mass, color, and SSFRs compared to those outside of filaments. 
\end{enumerate}

We conclude that the cosmic web filaments play an important role in the evolution of groups and their central galaxies. In the future, we plan to make quantitative comparisons of our observational results with hydrodynamical simulations to study in detail the cold streams of gas in filaments and their relationship to the properties of central galaxies. This will shed more light on the various physical mechanisms by which cosmic web filaments drive the evolution of galaxies. 

\begin{acknowledgements}
We thank the anonymous referee for constructive comments that improved the quality of the paper. AP acknowledges the financial support from Tuorla Observatory, University of Turku. We also acknowledge the support by the Estonian Research Council grants IUT26-2, IUT40-2, and by the Centre of Excellence “Dark side of the Universe” (TK133) financed by the European Union through the European Regional Development Fund. 

Funding for SDSS-III has been provided by the Alfred P. Sloan Foundation, the Participating Institutions, the National Science Foundation, and the U.S. Department of Energy Office of Science. The SDSS-III web site is \url{http://www.sdss3.org/}. 

SDSS-III is managed by the Astrophysical Research Consortium for the Participating Institutions of the SDSS-III Collaboration including the University of Arizona, the Brazilian Participation Group, Brookhaven National Laboratory, Carnegie Mellon University, University of Florida, the French Participation
Group, the German Participation Group, Harvard University, the Instituto de Astrofisica de Canarias, the Michigan State/Notre Dame/JINA Participation Group, Johns Hopkins University, Lawrence Berkeley National Laboratory,
Max Planck Institute for Astrophysics, Max Planck Institute for Extraterrestrial Physics, New Mexico State University, New York University, Ohio State University, Pennsylvania State University, University of Portsmouth, Princeton University, the Spanish Participation Group, University of Tokyo, University of Utah, Vanderbilt University, University of Virginia, University of Washington, and Yale University.
\end{acknowledgements}

\bibliographystyle{aa} 
\bibliography{references.bib} 
\onecolumn
\begin{appendix} 

\longtab[1]{
\section{Two sample Kolmogorov-Smirnov (K-S) tests}

We perform two sample Kolmogorov-Smirnov (K-S) test to determine whether the properties of galaxies or groups inside and outside of filaments are different or not at fixed large-scale luminosity density range. Using bootstrap resampling method, we draw randomly galaxies or groups from the two samples (inside and outside of filaments at fixed luminosity density) with repetitions and calculate mean properties separately. We repeat this process 1000 times, i.e. we have 1000 different estimates of mean properties inside and outside of filaments. We compare the resulting distributions of mean properties inside and outside of filaments using the K-S test. The results are summarized in Table~\ref{table:2}. The probability values of the properties tabulated are very low at fixed luminosity density range in majority of the mass bins. This means that the properties of galaxies or groups considered in this study are different inside and outside of filaments at fixed large-scale luminosity density. We also perform similar analysis to find out whether the properties of galaxies or groups inside and outside of filaments are different or not at fixed group masses. The results are summarized in Table~\ref{table:3}.
\begin{longtable}[!htbp] {|l|c|c|r|}

\caption{\label{table:2} Two sample K-S test results for comparison between properties of groups and their central galaxies inside and outside of filaments in different luminosity density ranges.}\\
\hline
\hline                        
Density ranges\tablefootmark{a}& Properties\tablefootmark{b}& Test statistic\tablefootmark{c} & Probability values\tablefootmark{d} \\    
\hline
\hline
\endfirsthead
\caption{Continued.} \\
\hline
Density ranges\tablefootmark{a}& Properties\tablefootmark{b}& Test statistic\tablefootmark{c} & Probability values\tablefootmark{d} \\    
\hline
\hline
\endhead
\hline
\endfoot                                
           &     $(M/L)_g$           & 0.694  & 2.2e-16 \\
		   &   Stellar mass       & 1.000  & 2.2e-16 \\
Den8 < 1.5 &      SSFR           & 0.898  & 2.2e-16 \\
           &  $g-r$ color         & 0.654  & 2.2e-16 \\
           &   Elliptical fraction         & 0.996  & 2.2e-16 \\
           &   Spiral fraction         & 0.708  & 2.2e-16 \\          
                     
\hline           
           &    $(M/L)_g$             & 0.416  & 2.2e-16 \\
 			& Stellar mass           & 1.000  & 2.2e-16 \\
 1.5 < Den8 < 5&       SSFR               & 0.750  & 2.2e-16 \\
           &   $g-r$ color            & 1.000  & 2.2e-16 \\
           &   Elliptical fraction         & 0.916  & 2.2e-16 \\          
           &   Spiral fraction         & 0.942  & 2.2e-16 \\          
           
           \hline
           &   $(M/L)_g$           & 0.880 & 2.2e-16 \\
		   &  Stellar mass            & 0.852 & 2.2e-16 \\
Den8 > 5   &         SSFR      		& 0.772 & 2.2e-16 \\           
           &   $g-r$ color         & 0.872  & 2.2e-16 \\          
           &   Elliptical fraction         & 0.956  & 2.2e-16 \\          
           &   Spiral fraction         & 0.686  & 2.2e-16 \\          
                                          
\hline                                             

\end{longtable}
\tablefoot{
\tablefoottext{a}{8 $h^{-1}$Mpc smoothed luminosity density ranges in units of cosmic mean density.}
\tablefoottext{b}{Properties of groups and central galaxies. $\phi(\log {M_{g}})$ denotes the number density of groups in each group mass bin. }
\tablefoottext{c}{K-S test statistics for distribution of mean properties of groups or central galaxies inside and outside of filaments.}
\tablefoottext{d}{Probability that the distribution of mean properties of groups or central galaxies inside and outside of filaments are similar.}

}
\begin{longtable} {|l|c|c|c|r|}

\caption{\label{table:3} Two sample K-S test results for comparison between properties of groups and their central galaxies inside and outside of filaments in different mass bins and luminosity density ranges.}\\
\hline
\hline                        
Density ranges\tablefootmark{a}& Properties\tablefootmark{b}& Mass bin\tablefootmark{c}  & Test statistic\tablefootmark{d} & Probability values\tablefootmark{e} \\    
                               &                            &  ($h^{-1}$M$_\odot$)             &  & \\
\hline
\hline
\endfirsthead
\caption{Continued.} \\
\hline
Density ranges\tablefootmark{a}& Properties\tablefootmark{b}& Mass bin\tablefootmark{c}  & Test statistic\tablefootmark{d} & Probability values\tablefootmark{e} \\    
                               &                            &  ($h^{-1}$M$_\odot$)             &  & \\
\hline
\hline
\endhead
\hline
\endfoot                                
           & 					 & 12--12.5 & 0.61  & 2.2e-16 \\
           &                      & 12.5--13 & 0.96  & 2.2e-16 \\
           &    $\phi(\log {M_{g}})$   & 13--13.5 & 0.25 & 5.362e-14 \\
           &                      & 13.5--14 & 0.55  & 2.2e-16 \\
           &                      & 14--14.5 & 0.596  & 2.2e-16 \\ 
           &                      & 14.5--15 & 0.84  & 2.2e-16 \\\cline{2-5} 

           & 		            & 12--12.5 & 0.586  & 2.2e-16 \\
           &                      & 12.5--13 & 0.374  & 2.2e-16 \\
Den8 > 5           &   $(M/L)_g$           & 13--13.5 & 0.368 & 2.2e-16 \\
           &                      & 13.5--14 & 0.766  & 2.2e-16 \\ 
           &                      & 14--14.5 & 0.502  & 2.2e-16 \\ 
           &                      & 14.5--15 & 0.424  & 2.2e-16 \\\cline{2-5} 
           & 		         & 12--12.5 & 0.236  & 1.61e-12 \\
           &                      & 12.5--13 & 0.252  & 3.242e-14 \\
           &  Stellar mass             		 & 13--13.5 & 0.968 & 2.2e-16 \\
           &               		 & 13.5--14 & 0.546  & 2.2e-16 \\
           &                      & 14--14.5 & 0.472  & 2.2e-16 \\ 
           &                      & 14.5--15 & 0.714  & 2.2e-16 \\\cline{2-5} 
           &    		        		 & 12--12.5 & 0.338  & 2.2e-16 \\
           &                      & 12.5--13 & 0.322  & 2.2e-16 \\
		   &         SSFR      		 & 13--13.5 & 0.882 & 2.2e-16 \\
           &               		 & 13.5--14 & 0.14  & 0.0001109 \\
           &                      & 14--14.5 & 0.41  & 2.2e-16 \\ 
           &                      & 14.5--15 & 0.472  & 2.2e-16 \\\cline{2-5}                
           &  			  		 & 12--12.5 & 0.238  & 1.002e-12 \\
           &               		 & 12.5--13 & 0.458  & 2.2e-16 \\
           &   $g-r$ color            		 & 13--13.5 & 0.37  & 2.2e-16 \\
           &               		 & 13.5--14 & 0.856  & 2.2e-16 \\
           &                      & 14--14.5 & 0.214  & 2.273e-10 \\ 
Den8 > 5           &                      & 14.5--15 & 0.672  & 2.2e-16 \\\cline{2-5}            

           &  			  		 & 12--12.5 & 0.562  & 2.2e-16 \\
           &               		 & 12.5--13 & 0.434  & 2.2e-16 \\
           &   Elliptical fraction & 13--13.5 & 0.966  & 2.2e-16 \\
           &               		 & 13.5--14 & 0.810  & 2.2e-16 \\
           &                      & 14--14.5 & 0.812  & 2.2e-16 \\ 
           &                      & 14.5--15 & 0.114  & 0.003013 \\\cline{2-5}
                                                                 
           &  			  		 & 12--12.5 & 0.188  & 4.228e-08 \\
           &               		 & 12.5--13 & 0.238  & 1.002e-12 \\
           &   Spiral fraction & 13--13.5 & 0.168  & 1.487e-06 \\
           &               		 & 13.5--14 & 0.276  & 2.2e-16 \\
           &                      & 14--14.5 & 0.316  & 2.2e-16 \\ 
           &                      & 14.5--15 & 0.516  & 2.2e-16 \\\cline{2-5}                                           
\hline
		   & 					 & 11.5--12 & 0.078 & 0.09547 \\      
           &                      & 12--12.5 & 0.93  & 2.2e-16 \\
           &     $\phi(\log {M_{g}})$                 & 12.5--13 & 0.638  & 2.2e-16 \\
           &                      & 13--13.5 & 0.864 & 2.2e-16 \\
           &                      & 13.5--14 & 0.418  & 2.2e-16 \\ 
           &                      & 14--14.5 & 0.132  & 0.0003292 \\\cline{2-5} 
           & 	            & 11.5--12 & 0.956 & 2.2e-16 \\      
           &                      & 12--12.5 & 0.646  & 2.2e-16 \\
		   &    $(M/L)_g$                  & 12.5--13 & 0.972  & 2.2e-16 \\
           &                      & 13--13.5 & 0.884 & 2.2e-16 \\
           &                      & 13.5--14 & 0.64  & 2.2e-16 \\ 
           &                      & 14--14.5 & 0.136  & 0.0001926 \\\cline{2-5} 
           & 			         & 11.5--12 & 0.376 & 2.2e-16 \\      
           &                  & 12--12.5 & 0.812  & 2.2e-16 \\
           &       Stellar mass  & 12.5--13 & 0.864  & 2.2e-16 \\
           &               		 & 13--13.5 & 0.962 & 2.2e-16 \\
           &               		 & 13.5--14 & 0.876  & 2.2e-16 \\
           &               		 & 14--14.5 & 0.19  & 2.897e-08 \\\cline{2-5}                 
1.5 < Den8 < 5  &           		 & 11.5--12 & 0.796 & 2.2e-16 \\      
           &               		 & 12--12.5 & 0.3   & 2.2e-16 \\
           &       SSFR              & 12.5--13 & 0.116  & 0.002394 \\
           &               		 & 13--13.5 & 0.744 & 2.2e-16 \\
           &               		 & 13.5--14 & 0.5  & 2.2e-16 \\
           &               		 & 14--14.5 & 0.388  & 2.2e-16 \\\cline{2-5}    
         
		   &  		    	         & 11.5--12 & 0.44 & 2.2e-16 \\      
           &               		 & 12--12.5 & 0.804  & 2.2e-16 \\
	       &   $g-r$ color            		 & 12.5--13 & 0.702  & 2.2e-16 \\
           &               		 & 13--13.5 & 0.99   & 2.2e-16 \\
           &               		 & 13.5--14 & 0.988  & 2.2e-16 \\ 
           &               		 & 14--14.5 & 0.284  & 2.2e-16 \\\cline{2-5} 

		   &  		    	         & 11.5--12 & 0.292 & 2.2e-16 \\      
           &               		 & 12--12.5 & 0.550  & 2.2e-16 \\
           &   Elliptical fraction& 12.5--13 & 0.100  & 0.01348 \\
           &               		 & 13--13.5 & 0.910   & 2.2e-16 \\
           &               		 & 13.5--14 & 0.140  & 0.0001109 \\ 
           &               		 & 14--14.5 & 0.228  & 1.03e-11 \\\cline{2-5} 

		   &  		    	         & 11.5--12 & 0.140 & 0.0001109 \\      
           &               		 & 12--12.5 & 0.096  & 0.01994 \\
1.5 < Den8 < 5           &   Spiral fraction   & 12.5--13 & 0.096  & 0.01994 \\
           &               		 & 13--13.5 & 0.208   & 8.061e-10 \\
           &               		 & 13.5--14 & 0.194  & 1.344e-08 \\ 
           &               		 & 14--14.5 & 0.102  & 0.01101 \\\cline{2-5} 
\hline
		   & 					 & 11.5--12 & 0.574 & 2.2e-16 \\      
           &                      & 12--12.5 & 0.09  & 0.03484 \\
           &     $\phi(\log {M_{g}})$& 12.5--13 & 0.376  & 2.2e-16 \\
           &                      & 13--13.5 & 0.788 & 2.2e-16 \\
           &                      & 13.5--14 & 0.928  & 2.2e-16 \\\cline{2-5}
           &                      & 11.5--12 & 0.284 & 2.2e-16 \\      
           &                      & 12--12.5 & 0.526  & 2.2e-16 \\
           &     $(M/L)_g$                 & 12.5--13 & 0.808  & 2.2e-16 \\
           &                      & 13--13.5 & 0.154 & 1.416e-05 \\
           &                      & 13.5--14 & 0.464  & 2.2e-16 \\\cline{2-5} 
	       & 			         & 11.5--12 & 0.226 & 1.622e-11 \\      
           &                      & 12--12.5 & 0.87  & 2.2e-16 \\
           &   Stellar mass       & 12.5--13 & 0.10  & 0.01348 \\
           &               		 & 13--13.5 & 0.798 & 2.2e-16 \\
           &               		 & 13.5--14 & 0.67  & 2.2e-16 \\\cline{2-5} 
		   &     	      		 & 11.5--12 & 0.168 & 1.487e-06 \\      
           &               		 & 12--12.5 & 0.398  & 2.2e-16 \\
Den8 < 1.5 &      SSFR           & 12.5--13 & 0.174  & 5.329e-07 \\
           &               		 & 13--13.5 & 0.874 & 2.2e-16 \\
           &               		 & 13.5--14 & 0.886  & 2.2e-16 \\\cline{2-5}
		   &      	 				& 11.5--12 & 0.194 & 1.344e-08 \\      
           &               		 & 12--12.5 & 0.554  & 2.2e-16 \\
           &  $g-r$ color             		 & 12.5--13 & 0.056  & 0.4131 \\
           &               		 & 13--13.5 & 0.828 & 2.2e-16 \\
           &               		 & 13.5--14 & 0.834  & 2.2e-16 \\\cline{2-5} 
           &      	 				& 11.5--12 & 0.626 & 2.2e-16 \\      
           &               		 & 12--12.5 & 0.266  & 2.2e-16 \\
           &  Elliptical fraction & 12.5--13 & 0.478  & 0.4131 \\
           &               		 & 13--13.5 & 0.878 & 2.2e-16 \\
           &               		 & 13.5--14 & 0.856  & 2.2e-16 \\\cline{2-5}  
           &      	 				& 11.5--12 & 0.276 & 2.2e-16 \\      
           &               		 & 12--12.5 & 0.114  & 2.2e-16 \\
           &  Spiral fraction & 12.5--13 & 0.08  & 0.4131 \\
           &               		 & 13--13.5 & 0.236 & 2.2e-16 \\
		 &  					 & 13.5--14 & 0.342  & 2.2e-16 \\\cline{2-5}
\hline
\end{longtable}
\tablefoot{
\tablefoottext{a}{8 $h^{-1}$Mpc smoothed luminosity density ranges in units of cosmic mean density.}
\tablefoottext{b}{Properties of groups and central galaxies. $\phi(\log {M_{g}})$ denotes the number density of groups in each group mass bin. }
\tablefoottext{c}{Group mass bins with width 0.5 $h^{-1}$M$_\odot$.}
\tablefoottext{d}{K-S test statistics for distribution of mean properties of groups or central galaxies inside and outside of filaments.}
\tablefoottext{e}{Probability that the distribution of mean properties of groups or central galaxies inside and outside of filaments are similar.}

}
}

\longtab[2]{

}
 

\end{appendix} 

\end{document}